\documentclass[prb,twocolumn,floatfix,showpacs]{revtex4-1}
\usepackage{graphicx,color}
\usepackage{amsmath,amssymb,bm}

\newcommand{\cprb}[3]{Phys.~Rev.~B {\bf #1}, #2 (#3)}
\newcommand{\cprl}[3]{Phys.~Rev.~Lett.~{\bf #1}, #2 (#3)}

\definecolor{darkred}{rgb}{0.90,0,0}
\definecolor{darkgreen}{rgb}{0,0.60,.2}
\definecolor{darkblue}{rgb}{0,0,1}
\definecolor{grey}{cmyk}{0,0,0,0.25}
\definecolor{orange}{cmyk}{0,0.6,0.8,0}

\newcommand{\ket}[1]{\left|#1\right\rangle}
\newcommand{\bra}[1]{\left\langle#1\right|}
\newcommand{\ii}{\text{i}}

\begin{document}
\title{\boldmath Dynamical phase transitions after quenches in non-integrable models}

\author{C.\ Karrasch$^{1,2}$ and D.\ Schuricht$^3$}

\affiliation{$^1$Department of Physics, University of California, Berkeley, CA 95720, USA}

\affiliation{$^2$Materials Sciences Division, Lawrence Berkeley National Laboratory, Berkeley, CA 94720, USA}

\affiliation{$^3$Institute for Theory of Statistical Physics, RWTH Aachen University and \\JARA--Fundamentals of Future Information Technology, 52056 Aachen, Germany}

\begin{abstract}
We investigate the dynamics following sudden quenches across quantum critical points belonging to  different universality classes. Specifically, we use matrix product state methods to study the quantum Ising chain in the presence of two additional terms which break integrability. We find that  in all models the rate function for the return probability to the initial state becomes a non-analytic function of time in the thermodynamic limit. This so-called `dynamical phase transition' was first observed in a recent work by Heyl, Polkovnikov, and Kehrein [Phys. Rev. Lett. \textbf{110}, 135704 (2013)] for the exactly-solvable quantum Ising chain, which can be mapped to free fermions. Our results for `interacting theories' indicate that non-analytic dynamics is a generic feature of sudden quenches across quantum critical points. We discuss potential connections to the dynamics of the order parameter.
\end{abstract}

\maketitle


\section{Introduction}

One of the central quantities in statistical mechanics\cite{Fisher,Mussardo} is the canonical partition function
\begin{equation}
Z(\beta)=\text{Tr}\bigl(e^{-\beta H}\bigr)=e^{-\beta L f(\beta)}~,
\label{eq:partitionfunction}
\end{equation}
where $\beta$ is the inverse temperature, $L$ denotes the system size, and $f(\beta)$ is the free energy density. A phase transition at a temperature $\beta_c$ is defined as a non-analytic point of the free energy density. For real temperatures and finite systems the partition function \eqref{eq:partitionfunction} is analytic, thus precluding the existence of a phase transition. In the thermodynamic limit, however, the free energy density may possess singularities. One way to characterize phase transitions was proposed by Yang and Lee.\cite{YangLee} The starting point\cite{Fisher,Thompson} is the observation that the partition function will have zeros in the complex $\beta$-plane. For a finite system these zeros are isolated and do not lie on  the real axis. For $L\to\infty$, however, the zeros may coalesce into lines which can cut the real axis at the critical temperature $\beta_c$. This picture was established in detail for the two-dimensional Ising model\cite{Fisher,2DIsing} as well as related systems. 

Recently, Heyl \emph{et al.}\cite{Heyl} pointed out the formal similarity between the partition function \eqref{eq:partitionfunction} and the return amplitude 
\begin{equation}
G(t)=\bra{\Psi_0}e^{-\ii Ht}\ket{\Psi_0}~.
\label{eq:overlapamplitude}
\end{equation}
For a quantum quench $G(t)$ is the Loschmidt amplitude,\cite{Silva08} i.e. the overlap of the initial state $\ket{\Psi_0}$ with its time evolution $e^{-\ii Ht}\ket{\Psi_0}$ under the post-quench Hamiltonian $H$. Specifically, Heyl \emph{et al.} studied the analytic properties of the boundary partition function
\begin{equation}
Z(z)=\bra{\Psi_0}e^{-zH}\ket{\Psi_0}
\label{eq:boundarypartitionfunction}
\end{equation}
as a function of $z\in\mathbb{C}$ and established close analogies with equilibrium phase transitions in statistical mechanics that we outlined above. For real $z=R$, Eq.~\eqref{eq:boundarypartitionfunction} can be interpreted~\cite{LeClair-95,Mussardo} as the partition function of  a system of length $R$ with boundary conditions described by the boundary state $\ket{\Psi_0}$; for $z=\ii t$ one recovers the return amplitude of Eq.~\eqref{eq:overlapamplitude}. 

Heyl \emph{et al.}\cite{Heyl} specifically investigated the prototypical quantum Ising chain which exhibits a quantum phase transition\cite{Sachdev} between ferromagnetic and paramagnetic ground states. The Ising chain can be mapped to a model of free fermions and thus allows an exact evaluation of the boundary partition function.\cite{LeClair-95,Silva08} If $|\Psi_0\rangle$ and $H$ are associated with the same phase, the lines of zeros of Eq.~\eqref{eq:boundarypartitionfunction} do not intersect the imaginary axis, while for a quench across the quantum phase transition zeros with $z=\ii t$ exist. The vanishing overlap of the time-evolved and the initial state translates into non-analyticities of the rate function for the return probability $Ll(t)=-\ln|G(t)|^2$ and can thus be viewed as a \emph{dynamical phase transition}.\cite{Heyl} Similarly to its thermal counterpart, the non-analyticities in time \emph{only appear in the thermodynamic limit}. They lead to a breakdown of short-time expansions analogously to the breakdown of high-temperature expansions in the vicinity of thermal critical points.\cite{Mitra} The characteristic times at which the non-analyticities in the rate functions for the return amplitude and probability occur were also observed in the dynamics of the order parameter.\cite{Heyl,Calabrese-12-2} 

In this paper we address the following question: Is a dynamical phase transition a generic feature of quantum quenches across quantum critical points? According to our numerical results, the answer is clearly \emph{yes}. From the study of the quantum Ising chain this is not \emph{a priori} clear, as the Ising chain can be mapped to an \emph{integrable} model of \emph{free fermions} whose dynamics might be special.

To be more precise, we use the time-dependent density-matrix renormalization group (DMRG) to investigate quenches in the quantum Ising chain (where the comparison with the exact solution allows to test our numerics), the transverse axial next-nearest-neighbour Ising (ANNNI) model (which corresponds to the Ising model complemented by additional next-nearest neighbor interactions), and a generalized quantum Ising chain in a tilted magnetic field. While the quantum critical points in the first two models are of Ising type, in the latter model it belongs to a different universality class. Furthermore, the ANNNI model as well as the generalized Ising chain cannot be mapped to a non-interacting theory and are, to the best of our knowledge, not integrable. In all three models we consistently observe that when quenching across the quantum critical point the rate function for the return probability shows non-analyticities at critical times $t^*_n$, while for quenches within the same phase the time evolution of the rate function is completely smooth. Our results for `interacting theories' thus indicate that non-analytic dynamics is a generic feature of quenches across critical points independently of their universality class. We also discuss potential connections to the dynamics of the order parameter.

This exposition is organized as follows: In Sec.~\ref{sec:Ising} we first review the results of Ref.~\onlinecite{Heyl} for the quantum Ising chain. We test our DMRG numerics against the exact solution and present new data for a quench starting out of a spin-polarized ground state. In Sec.~\ref{sec:ANNNI} we discuss DMRG results for the ANNNI model. The behavior of the return amplitude is very similar to the quantum Ising chain -- we observe non-analyticities even if the model is strongly non-integrable. In Sec.~\ref{sec:genIsing} we study a generalized quantum Ising chain whose quantum phase transition does not belong to the Ising universality class. For this model we also find non-analytic behavior of the return amplitude, although for certain parameters the non-analyticity may appear only at late times. We summarize our results in Sec.~\ref{sec:conclusion}. Technical details including the implementation of the DMRG calculations are elaborated on in the appendix.

\section{Quantum Ising chain}\label{sec:Ising}

\begin{figure}[t]
\includegraphics[width=0.95\linewidth,clip]{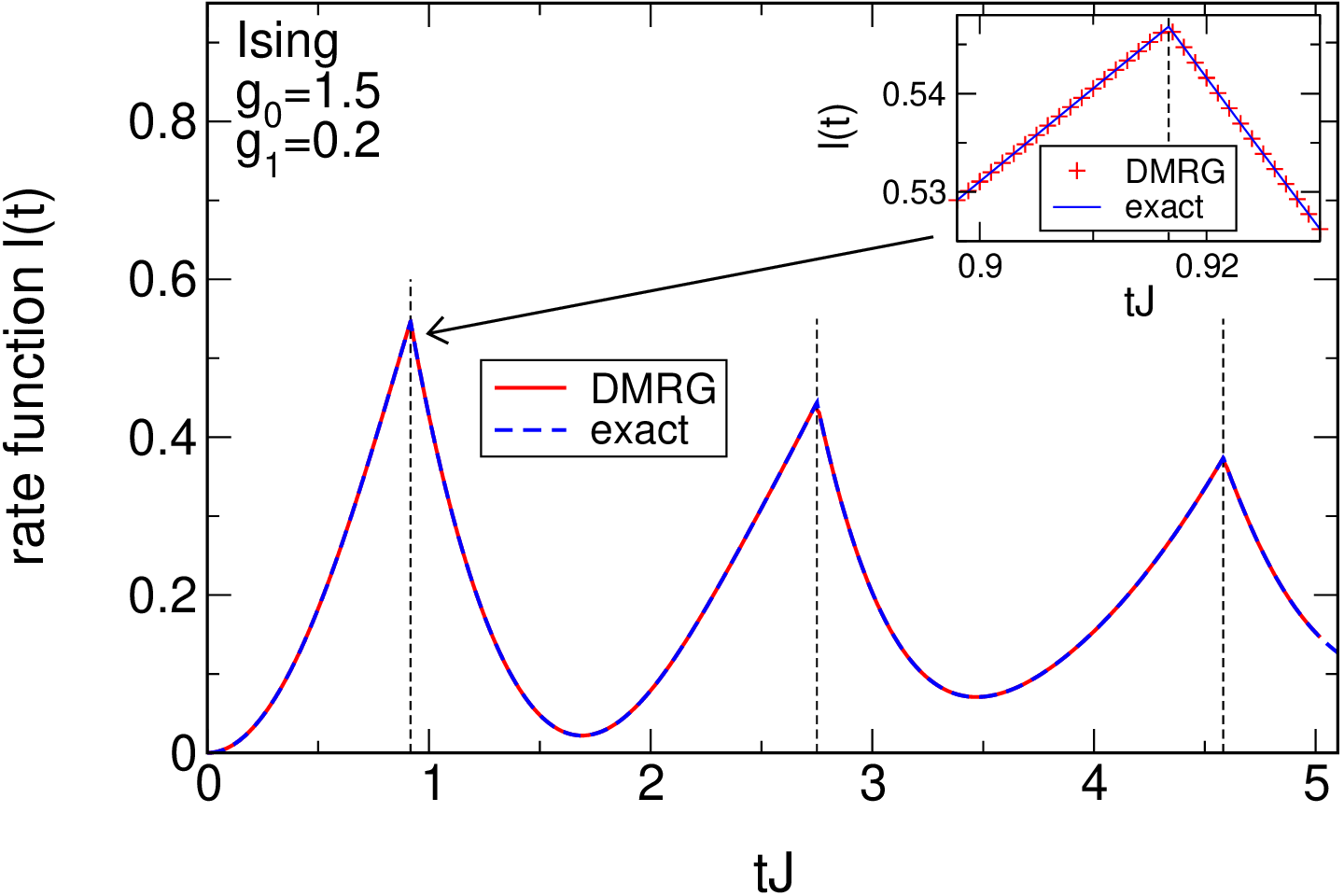}
\caption{(Color online) Rate function $l(t)$ defined in Eq.~\eqref{eq:returnamplitude} for a quench from the PM to the FM phase in the Ising model of Eq.~\eqref{eq:Ising}. $l(t)$ characterizes the overlap with the initial state in the thermodynamic limit. The DMRG data agree well with the analytic result obtained by mapping the model to free fermions [Eq.~\eqref{eq:Isingfermion}; see also Ref.~\onlinecite{Heyl}]. The rate function shows non-analytic behavior at the times $t_n^*$ indicated by the vertical dashed lines [Eq.~\eqref{eq:tnstar}].}
\label{fig:ising1}
\end{figure}

\begin{figure*}[t]
\includegraphics[width=0.475\linewidth,clip]{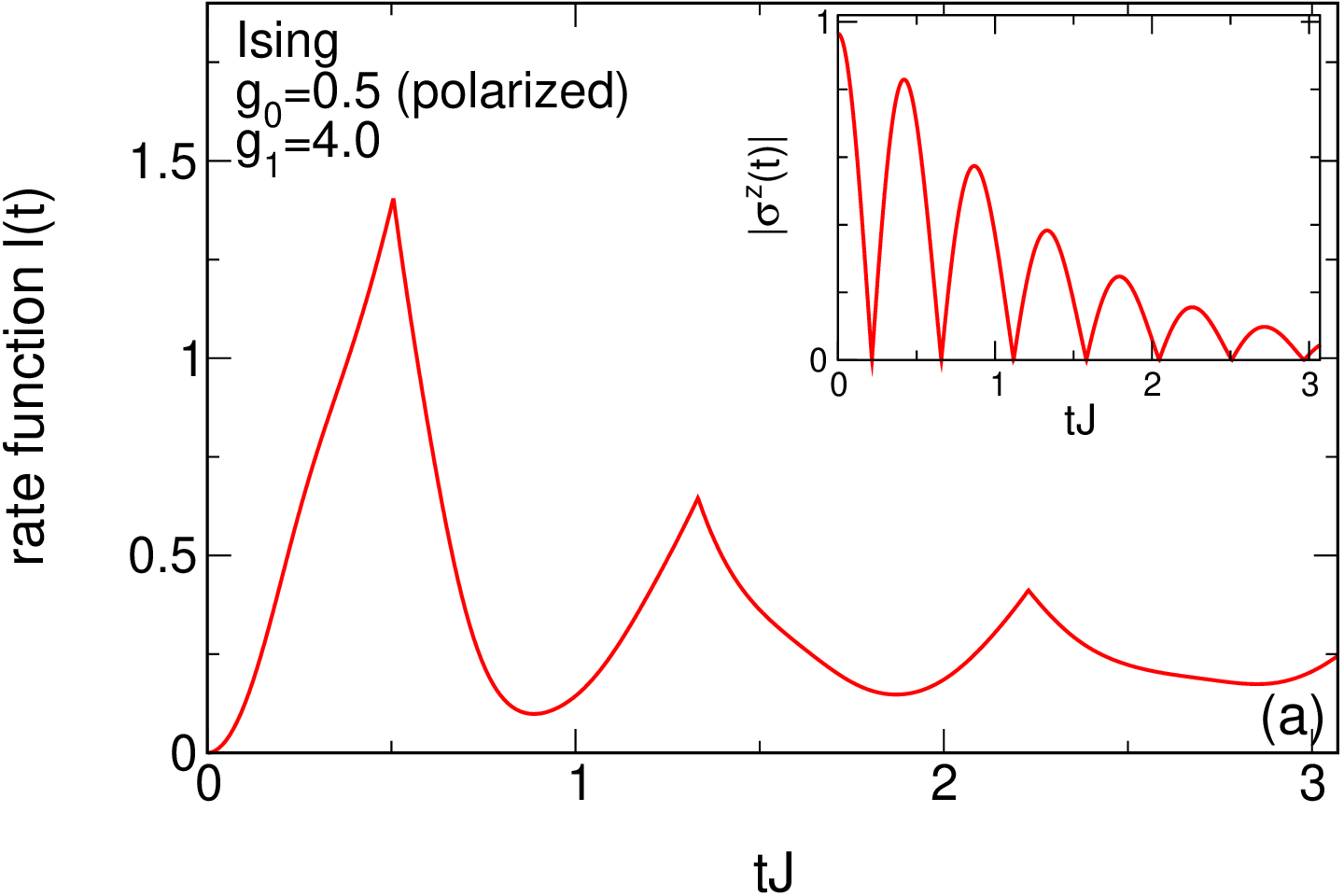}
\hspace*{0.025\linewidth}
\includegraphics[width=0.475\linewidth,clip]{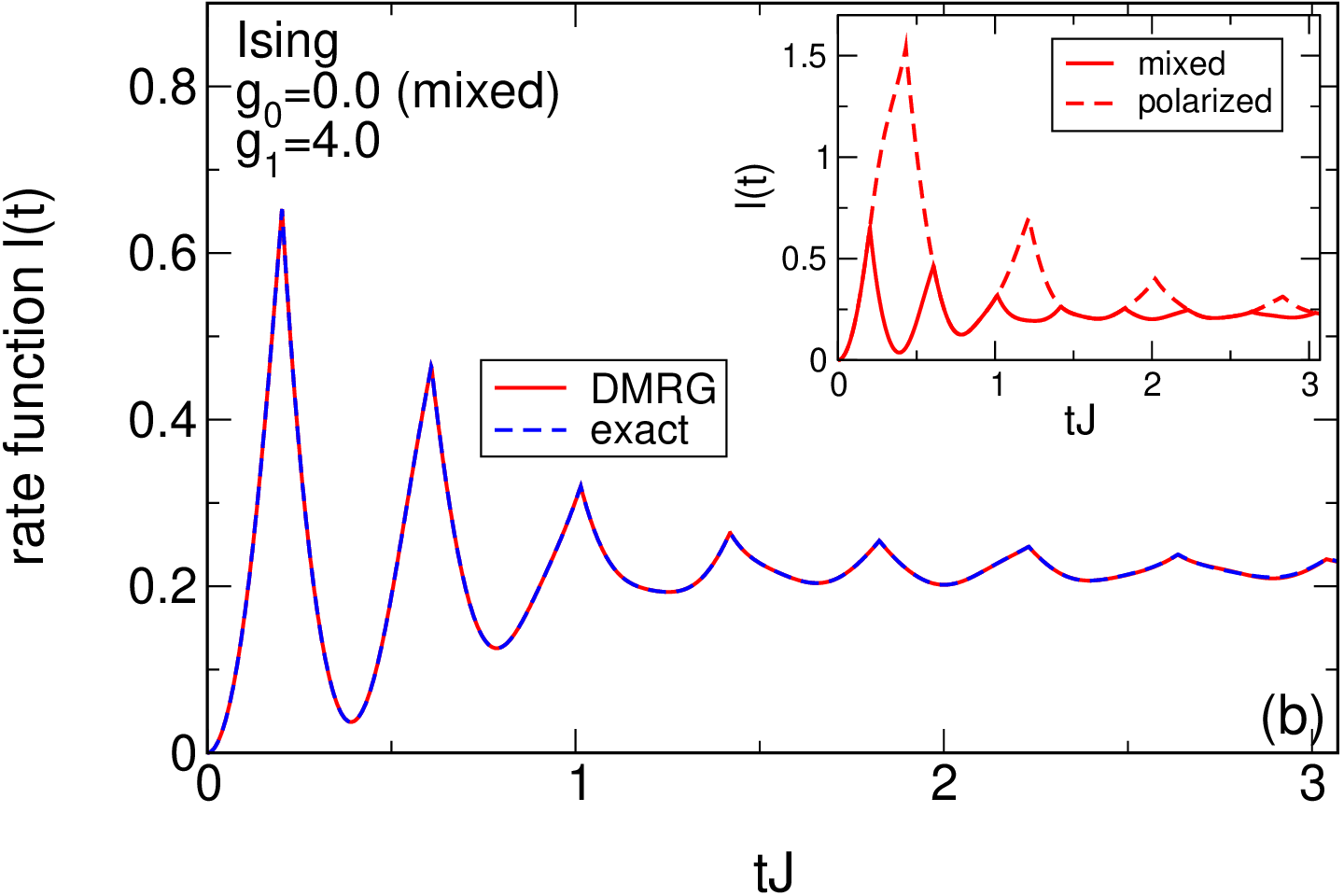}
\caption{(Color online) The same as in Fig.~\ref{fig:ising1} but for quenches from the FM to the PM phase in the Ising model. In the thermodynamic limit, the ground state $|\pm\rangle$ within the FM phase is two-fold degenerate. (a) Quench performed starting from the polarized state $\ket{+}$. The rate function shows non-analytic behavior which, however, does not occur at the times $t_n^*$ defined in Eq.~\eqref{eq:tnstar}. Inset: Time evolution of the order parameter $\langle\sigma^z_i(t)\rangle$, which at sufficiently late times oscillates with the frequency $2t^*$. (b) Quench starting from the mixed state $\ket{\text{NS}}=(\ket{+}-\ket{-})/\sqrt{2}$. The DMRG data agree well with the analytic result obtained for the corresponding quench in the fermionic model\cite{Heyl} of Eq.~\eqref{eq:Isingfermion}. The rate function shows non-analytic behavior at the times $t_n^*$. Inset: Comparison of the rate functions starting from the mixed and polarized states $\ket{\text{NS}}$ and $\ket{+}$, respectively.}
\label{fig:ising2}
\end{figure*}

We begin our study with the quantum Ising chain
\begin{equation}
H_\text{Ising}=-J\sum_i\bigl(\sigma_i^z\sigma_{i+1}^z+g\sigma_i^x\bigr)~,
\label{eq:Ising}
\end{equation}
where we assume $J>0$ and $g\ge 0$. The model exhibits a quantum phase transition\cite{Sachdev} at $g_c=1$, which separates a ferromagnetic (FM) phase for $g<g_c$ from a paramagnetic (PM) phase for $g>g_c$. In the FM the system possesses two degenerate ground states $\ket{\pm}$ with $\langle\sigma_i^z\rangle\neq 0$, while the PM ground state with $\langle\sigma_i^z\rangle=0$ is unique. Close to the quantum critical point $g=g_c$ the correlation length $\xi$ diverges as $\xi\sim|g-g_c|^{-\nu}$ with $\nu=1$. Quenches across this quantum phase transition have been studied by various authors.\cite{Silva08,Calabrese-12-2,Pollmann-10,Isingquenches} In particular, Heyl \emph{et al.}\cite{Heyl} investigated the analytic properties of the boundary partition function \eqref{eq:boundarypartitionfunction} and showed that its lines of zeros cut the imaginary time axis $z=\ii t$ for quenches across the quantum phase transition, giving rise to a dynamical phase transition at a critical time $t^*/2$ [see Eq.~\eqref{eq:tstar} below].

It is well-known that the quantum Ising chain can be mapped to a model of non-interacting fermions via a Jordan-Wigner transformation (see e.g. Ref.~\onlinecite{Calabrese-12-2}), and is thus certainly integrable.\cite{CauxMossel11} In the fermionic languange, the Hamiltonian can be easily diagonalized, 
\begin{equation}
H_\text{Ising}=\sum_k\epsilon_g(k)\,\alpha_k^\dagger\alpha_k~,
\label{eq:Isingfermion}
\end{equation}
where
\begin{equation}
\epsilon_g(k)=2J\sqrt{(g-\cos k)^2+\sin^2 k}~,
\label{eq:Isingenergy}
\end{equation}
and $\alpha_k^\dagger$ and $\alpha_k$ are fermionic creation and annihilation operators. The fermions fulfill either anti-periodic or periodic boundary conditions, usually referred to as Neveu-Schwarz (NS) or  Ramond (R) sectors, respectively. The momenta $k$ are correspondingly quantized as either half-integer or integer multiples of $2\pi/L$. Clearly, at $g=g_c=1$ the dispersion in Eq.~\eqref{eq:Isingenergy} becomes gapless, indicating the existence of the quantum phase transition. We note that the mapping from the spin model \eqref{eq:Ising} to the fermionic model \eqref{eq:Isingfermion} involves a non-local transformation of the spin operators, which leads to subtleties for quenches originating in the FM (see below).

The quench protocol in the Ising model is implemented by suddenly switching the transverse field from its initial value $g=g_0$ to its final value $g=g_1$. The boundary partition function for the equivalent quench in the fermionic model can be calculated explicitly\cite{Silva08,Heyl} and, up to a constant, in the thermodynamic limit reads
\begin{equation}
Z(z)=\exp\bigl[-Lf(z)\bigr]~,
\label{eq:partitionfunctionIsing}
\end{equation}
with 
\begin{equation}
f(z)=-\int_0^\pi\frac{\text{d}k}{2\pi}\,\ln\bigl[\cos^2\phi_k+\sin^2\phi_k\,e^{-2z\epsilon_{g_1}(k)}\bigr]~,
\label{eq:freeenergydensity}
\end{equation}
where $\phi_k=\theta_{g_0}(k)-\theta_{g_1}(k)$ and $\theta_g(k)=\arctan[\sin k/(g-\cos k)]/2\in[0,\pi/2]$. We stress that Eq.~\eqref{eq:partitionfunctionIsing} only applies to quenches starting from the unique ground state of the fermionic model, which lies in the NS sector for any finite system. In the PM phase of the Ising model this state corresponds to the unique ground state, while in the FM phase it corresponds to a superposition of the degenerate ground states $\ket{\pm}$.\cite{Calabrese-12-2} Hence, quenches originating in one of the spin-polarized states $\ket{\pm}$ are not described by Eq.~\eqref{eq:freeenergydensity}. 

The zeros of Eq.~\eqref{eq:partitionfunctionIsing} are located along the lines\cite{Heyl}
\begin{equation}
z_n(k)=\frac{1}{2\epsilon_{g_1}(k)}\bigl[\ln(\tan^2\phi_k)+\ii\pi(2n+1)\bigr],\; n\in\mathbb{Z}~,
\end{equation}
where the argument of the logarithm in Eq.~\eqref{eq:freeenergydensity} vanishes. For quenches across the quantum critical point these lines of zeros cross the imaginary time axis, that is there exist zeros with $\mathfrak{Re}\,z_n(k^*)=0$ for $k^*=\text{arccos}[(1+g_0g_1)/(g_0+g_1)]$. This leads to a non-analyticity in the rate function for the return probability
\begin{equation}
l(t)=-\frac{1}{L}\ln\big|G(t)\big|^2=2\,\mathfrak{Re}\,f(\ii t)
\label{eq:returnamplitude}
\end{equation}
at the times
\begin{equation}
t_n^*=\mathfrak{Im}\,z_n(k^*)=t^*\left(n+\frac{1}{2}\right)~,
\label{eq:tnstar}
\end{equation}
with
\begin{equation}
t^*=\frac{\pi}{\epsilon_{g_1}(k^*)}=\frac{\pi}{2J}\sqrt{\frac{g_0+g_1}{(g_0-g_1)(1-g_1^2)}}~.
\label{eq:tstar}
\end{equation}
We note that either $g_0<1$ and $g_1>1$ or vice versa. Eq.~\eqref{eq:tstar} represents a new time scale generated by the quench. Eq.~\eqref{eq:tstar} remains finite in the vicinity of the quantum critical point, where the system can be described\cite{Mussardo} in terms of massive Majorana fields (see App.~\ref{sec:appsl}). This time scale was also observed~\cite{Heyl,Calabrese-12-2} as an oscillation frequency in the time evolution of the order parameter $\langle\sigma^z_i\rangle$ for quenches from the FM to the PM when starting in one of the polarized states $\ket{\pm}$.

After this brief review we turn now to our DMRG calculations for quenches across the quantum critical point and show how the results discussed above can be recovered using this approach. 

\subsection{DMRG results: Quench PM to FM}

In this section we discuss time-dependent DMRG data for quenches across the quantum critical point in the Ising model \eqref{eq:Ising}. We start with the simpler quench from the PM to the FM and discuss the more subtle quench from the FM to the PM in the next subsection. Unless stated otherwise, all DMRG calculations were carried out directly in the thermodynamic limit $L\to\infty$ using an infinite-system algorithm. Details of the numerical implementation can be found in App.~\ref{sec:appDMRG}.

As discussed above, the ground state in the PM phase of the Ising model corresponds to the ground state of the fermionic model \eqref{eq:Isingfermion}. Thus, Eq.~\eqref{eq:freeenergydensity} is applicable, and the rate function \eqref{eq:returnamplitude} can be directly obtained. The comparison to the DMRG data is shown in Fig.~\ref{fig:ising1}; the agreement is good. In particular, we observe non-analyticities at the times $t_n^*$ defined in Eq.~\eqref{eq:tnstar}. In contrast, for quenches within one phase the rate function is a smooth function without features at the times $t_n^*$ [we show an example for this behavior for the ANNNI model in Fig.~\ref{fig:ANNNI1}(b)].

Further insight can be gained by considering the quench from $g_0=\infty$ to $g_1=0$. For this quench the initial state is simply given by $\ket{\rightarrow\rightarrow\ldots\rightarrow}$ where $\ket{\rightarrow}=(\ket{\uparrow}+\ket{\downarrow})/\sqrt{2}$, while the time evolution is governed by the classical Hamiltonian $H=-J\sum_i\sigma_i^z\sigma_{i+1}^z$. Since all terms in $H$ commute, the rate function of Eq.~\eqref{eq:returnamplitude} can be easily calculated and is given by (if $L$ is divisible by four)
\begin{equation}
l_x(t)=-\frac{2}{L}\ln\bigl[\cos^{L}(Jt)+\sin^{L}(Jt)\bigr]~.
\label{eq:returnamplitudetrivialPMtoFM}
\end{equation}
This function has maxima at $t=t_n^*$ with $t^*=\pi/(2J)$, at which it becomes non-analytic in the thermodynamic limit $L\to\infty$. Thus we explicitly see that the thermodynamic limit is essential for the appearance of dynamical phase transitions.

\subsection{DMRG results: Quench FM to PM}

As a next step we consider a quench originating in the FM phase. We assume that the system is prepared in the state $\ket{+}$ with $\langle\sigma^z_i\rangle>0$ and subsequently quench to the PM phase with $g_1>1$. The corresponding rate function is shown in Fig.~\ref{fig:ising2}(a). We clearly observe non-analyticities. However, $l(t)$ cannot be described using the analytic result of Eq.~\eqref{eq:freeenergydensity}, and the non-analyticities no longer occur at the times $t_n^*$. This can be attributed to the fact that the state $\ket{+}$ does not correspond to the ground state of the fermionic model \eqref{eq:Isingfermion}; hence, Eq.~\eqref{eq:freeenergydensity} is not applicable. 

One can gain some analytic understanding by considering the simple (but instructive) quench from $g_0=0$ to $g_1=\infty$, where the time evolution after the quench is governed by the trivial Hamiltonian $H'=-Jg_1\sum_i\sigma_i^x$ (see also supplementary material to Ref.~\onlinecite{Heyl}). The initial state for $g_0=0$ is given by $\ket{+}=\ket{\uparrow\uparrow\ldots\uparrow}$, and a straightforward calculation yields
\begin{equation}
l_+(t)=-2\ln\big|\cos(Jg_1t)\big|~.
\label{eq:returnamplitudetrivialquench}
\end{equation}
Thus, non-analyticities occur at the times $\tilde{t}_n=2t^*(n+1/2)$, $t^*=\pi/(2Jg_1)$, which differ from the times $t_m^*$ for all $m,n$. Since the spins in $H'$ do not interact at all, the rate function is perfectly periodic and does not decay. For general quenches starting from a spin-polarized state, such a periodicity cannot be observed [see Fig.~\ref{fig:ising2}(a)]. Another consequence of the decoupled time evolution of individual spins is that the non-analytic behavior shows up also in finite systems. In contrast, for quenches to $g_1<\infty$ the time evolution is no longer trivial and $l_+(t)$ becomes smooth for finite systems.

The inset to Fig.~\ref{fig:ising2}(a) shows the time evolution of the order parameter $\langle\sigma^z_i(t)\rangle$. As reported previously,\cite{Heyl,Calabrese-12-2} at sufficiently late times we observe oscillations with the frequency $2t^*$. For the simple quench discussed above we specifically obtain $\langle\sigma^z_i(t)\rangle=\cos(2Jg_1t)$. We note, however, that in the general case the relation between the times of non-analytic behavior in the rate function and the oscillation frequency of the order parameter is unclear (see the discussion for the ANNNI model below). 

In order to make contact to the analytic result of Eq.~\eqref{eq:freeenergydensity} also for quenches starting in the FM phase, we note that the ground state of the fermionic model \eqref{eq:Isingfermion} corresponds to  a superposition of the FM ground states\cite{Calabrese-12-2} $\ket{\text{NS}}=(\ket{+}-\ket{-})/\sqrt{2}$. DMRG data for a quench starting from this state is shown in Fig.~\ref{fig:ising2}(b); the agreement with the analytic result is again good. In particular, the rate function possesses non-analyticities at the times $t_n^*$. As the initial state is an equal superposition of $\ket{+}$ and $\ket{-}$, the order parameter $\langle\sigma^z_i(t)\rangle$ vanishes identically at all times. 

In the inset to Fig.~\ref{fig:ising2}(b) we compare the rate functions for quenches starting from the polarized state $\ket{+}$ and the mixed state $\ket{\text{NS}}$. We observe that for half of the time the rate functions are identical. This can again be understood by considering the simple quench from $g_0=0$ to $g_1=\infty$ introduced above. One finds that (if $L$ is divisible by four)
\begin{equation}
l_\text{NS}(t)=-\frac{2}{L}\ln\bigl[\cos^{L}(Jg_1t)+\sin^{L}(Jg_1t)\bigr]~.
\label{eq:returnamplitudetrivialquenchNS}
\end{equation}
For large systems this shows a switching behavior\cite{Heyl} depending on whether the first or the second term in the argument of the logarithm dominates, in complete analogy with the DMRG data for generic quenches from the FM to the PM. The non-analyticities of Eq.~\eqref{eq:returnamplitudetrivialquenchNS} follow in analogy to the ones of $l_x(t)$ at $t=t^*_n$ with $t^*=\pi/(2Jg_1)$.

\begin{figure}[b] \centering
\includegraphics[width=0.95\linewidth,clip=true]{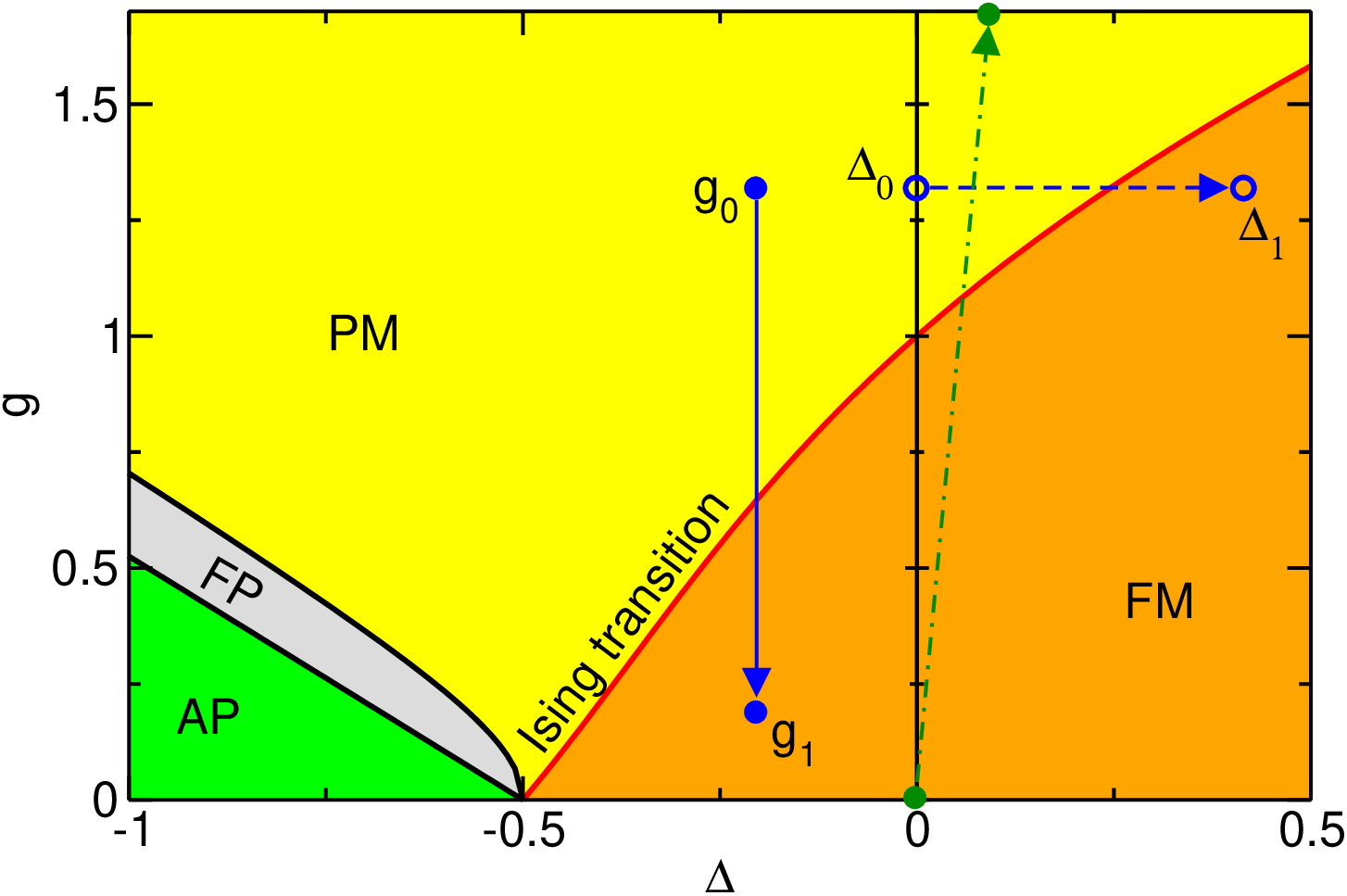}
\caption{(Color online) Sketch of the phase diagram~\cite{ANNNI,ANNNI2} of the ANNNI model defined in Eq.~\eqref{eq:ANNNI} as a function of $\Delta$ and $g$. There are four phases: a paramagnetic (PM) phase, a ferromagnetic (FM) phase, an anti phase (AP), and a floating phase (FP). The PM and FM phases are separated by an Ising transition located at $g_c(\Delta)$ defined in Eq.~\eqref{eq:ANNNIphasetransition}. We study quenches across the Ising transition as indicated by the solid arrow [see Fig.~\ref{fig:ANNNI1}(a)], the dashed arrow [see Fig.~\ref{fig:ANNNI1}(b)], and the dashed-dotted arrow (see Fig.~\ref{fig:ANNNI2}).}
\label{fig:phasediagramANNNI}
\end{figure} 

\begin{figure*}[t]
\includegraphics[width=0.475\linewidth,clip]{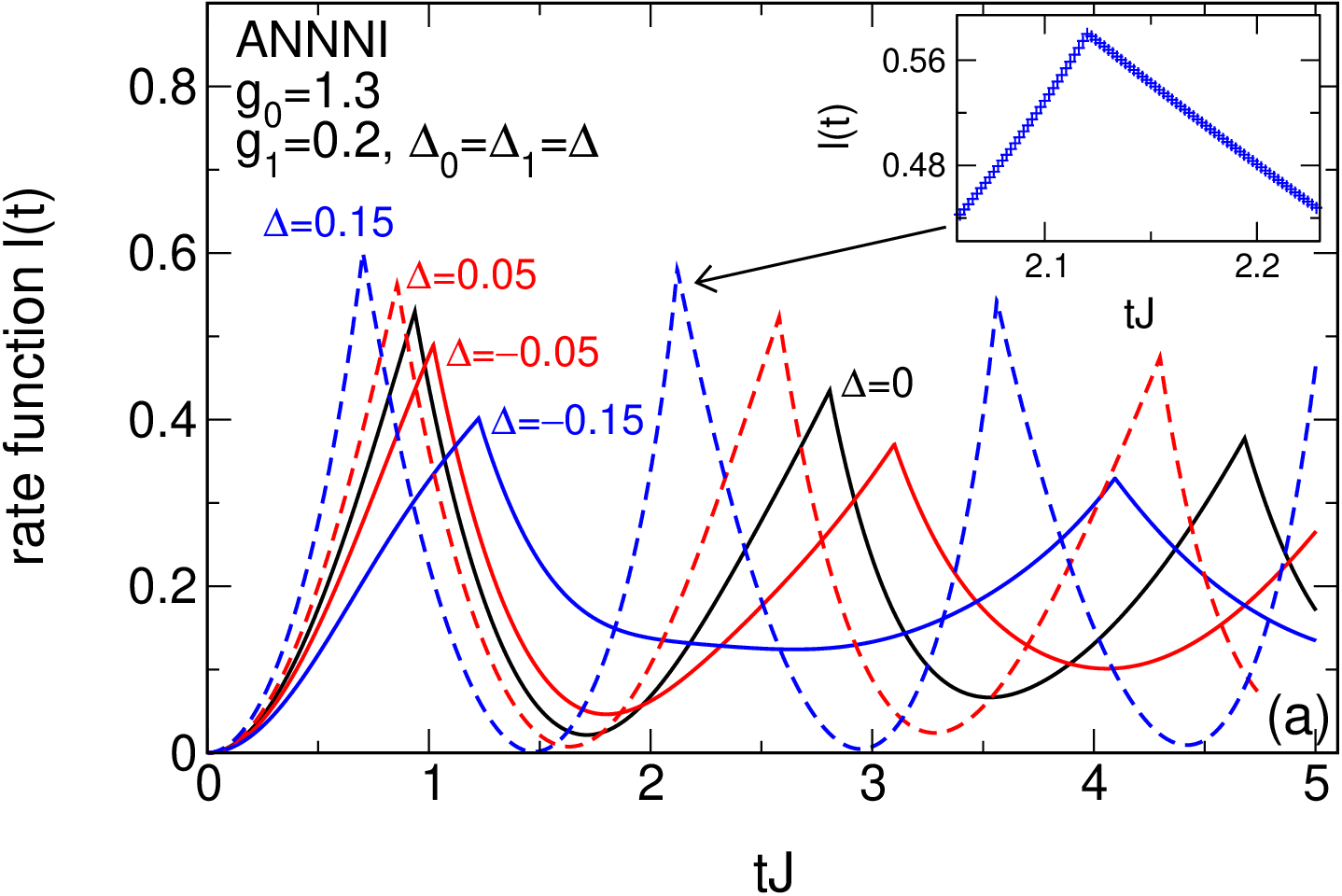}
\hspace*{0.025\linewidth}
\includegraphics[width=0.475\linewidth,clip]{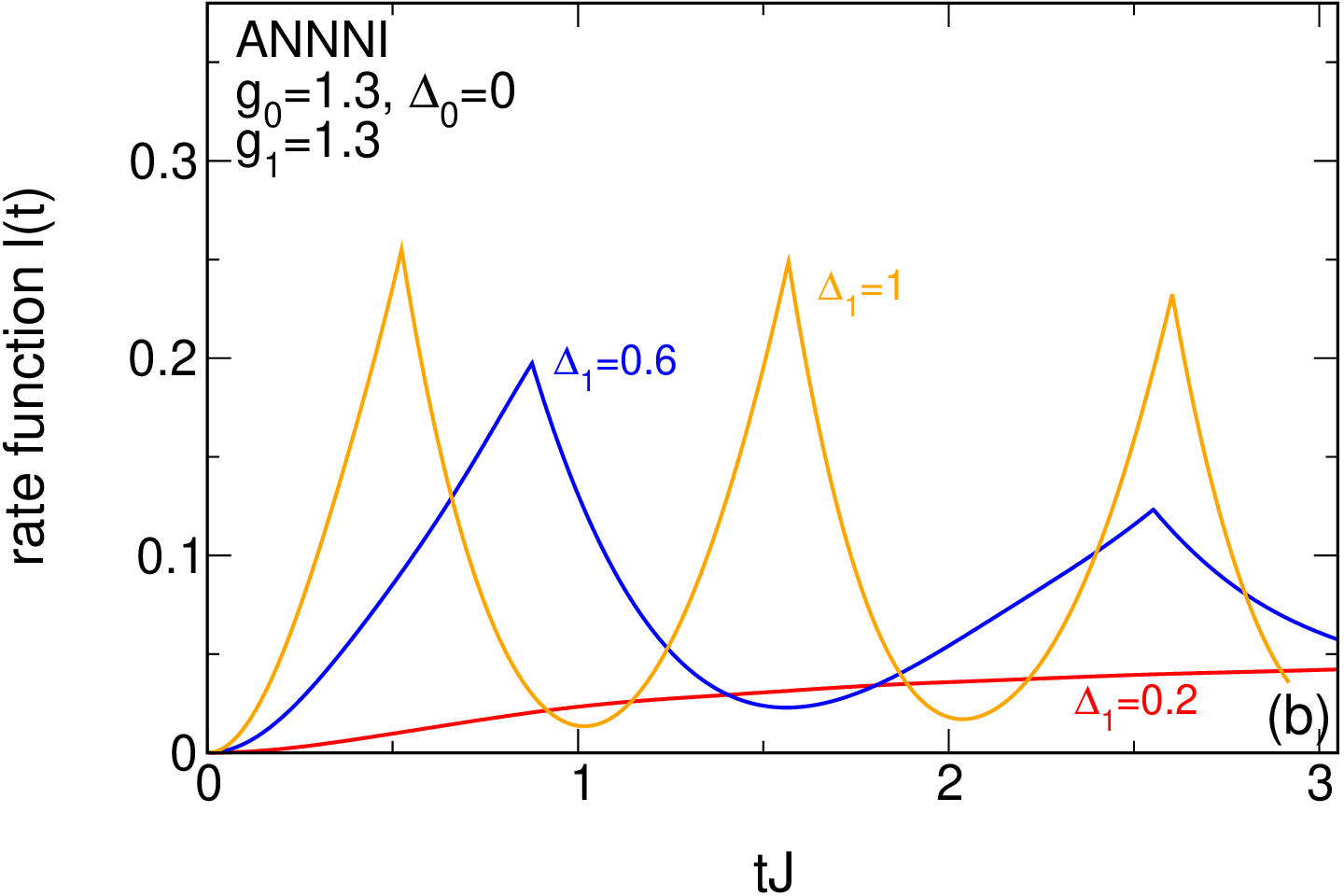}
\caption{(Color online) Rate function for a quench from the PM to the FM phase of the quantum Ising model in presence of integrability-breaking next-nearest neighbor interactions [the so-called ANNNI model; see Eq.~\eqref{eq:ANNNI}]. (a) Quench in the transverse field $g$ (indicated by the solid arrow in the phase diagram shown in Fig.~\ref{fig:phasediagramANNNI}). (b) Quench in the next-nearest neighbor interaction $\Delta$ (dashed arrow in Fig.~\ref{fig:phasediagramANNNI}). In complete analogy with the integrable `non-interacting' Ising chain ($\Delta=0$), the rate function exhibits non-analytic behavior as a function of time in the thermodynamic limit if one quenches across a critical point [note that the curve in (b) for $\Delta_1=0.2$ corresponds to a quench \textit{within} the PM phase].}
\label{fig:ANNNI1}
\end{figure*}

With this we conclude our analysis of quenches in the Ising model. In the remainder of this paper we address the question whether non-analytic behavior in the rate function for the return probability can be observed for quenches across quantum critical points in other models. We begin by considering the ANNNI model in the following section. 

\section{ANNNI model}\label{sec:ANNNI}

As second model we investigate the transverse axial next-nearest-neighbour Ising (ANNNI) model\cite{ANNNI} defined by the Hamiltonian
\begin{equation}
H_\text{ANNNI}=-J\sum_i\bigl[\sigma_i^z\sigma_{i+1}^z+\Delta\sigma_i^z\sigma_{i+2}^z+g\sigma_i^x\bigr]~.\label{eq:ANNNI}
\end{equation}
Again we assume $J>0$ and $g\ge 0$, while $\Delta$ can be positive or negative. Obviously, for $\Delta=0$ we recover the quantum Ising chain of Eq.~\eqref{eq:Ising}. We note that Eq.~\eqref{eq:ANNNI} is invariant under $g\to -g$ due to the transformation $\sigma_i^{x,z}\to-\sigma_i^{x,z}$. Using a Jordan-Wigner transformation, the ANNNI model can be mapped to a model of \emph{interacting} fermions. To the best of our knowledge, the resulting system is not integrable and does not allow an exact solution like the quantum Ising chain.

The phase diagram of the ANNNI model contains four phases (see Fig.~\ref{fig:phasediagramANNNI}):\cite{ANNNI,ANNNI2} A paramagnetic phase (PM) with a unique ground state satisfying $\langle\sigma_i^z\rangle=0$; a ferromagnetic phase (FM) with doubly degenerate ground state with $\langle\sigma_i^z\rangle\neq 0$; an ``anti phase" (AP) that schematically looks like $\uparrow\uparrow\downarrow\downarrow\uparrow\uparrow\downarrow\downarrow$; and a ``floating phase" (FP) between the PM and the AP. The phase transition between the PM and the FM is in the Ising universality class with $\nu=1$. For $\Delta<0$ it is located at 
\begin{equation}
1+2\Delta=g_c+\frac{\Delta g_c^2}{2(1+\Delta)}~.
\label{eq:ANNNIphasetransition}
\end{equation}
We will restrict ourselves to quenches across this phase transition in the following (see the arrows in Fig.~\ref{fig:phasediagramANNNI}). 

\begin{figure}[b]
\includegraphics[width=0.95\linewidth,clip]{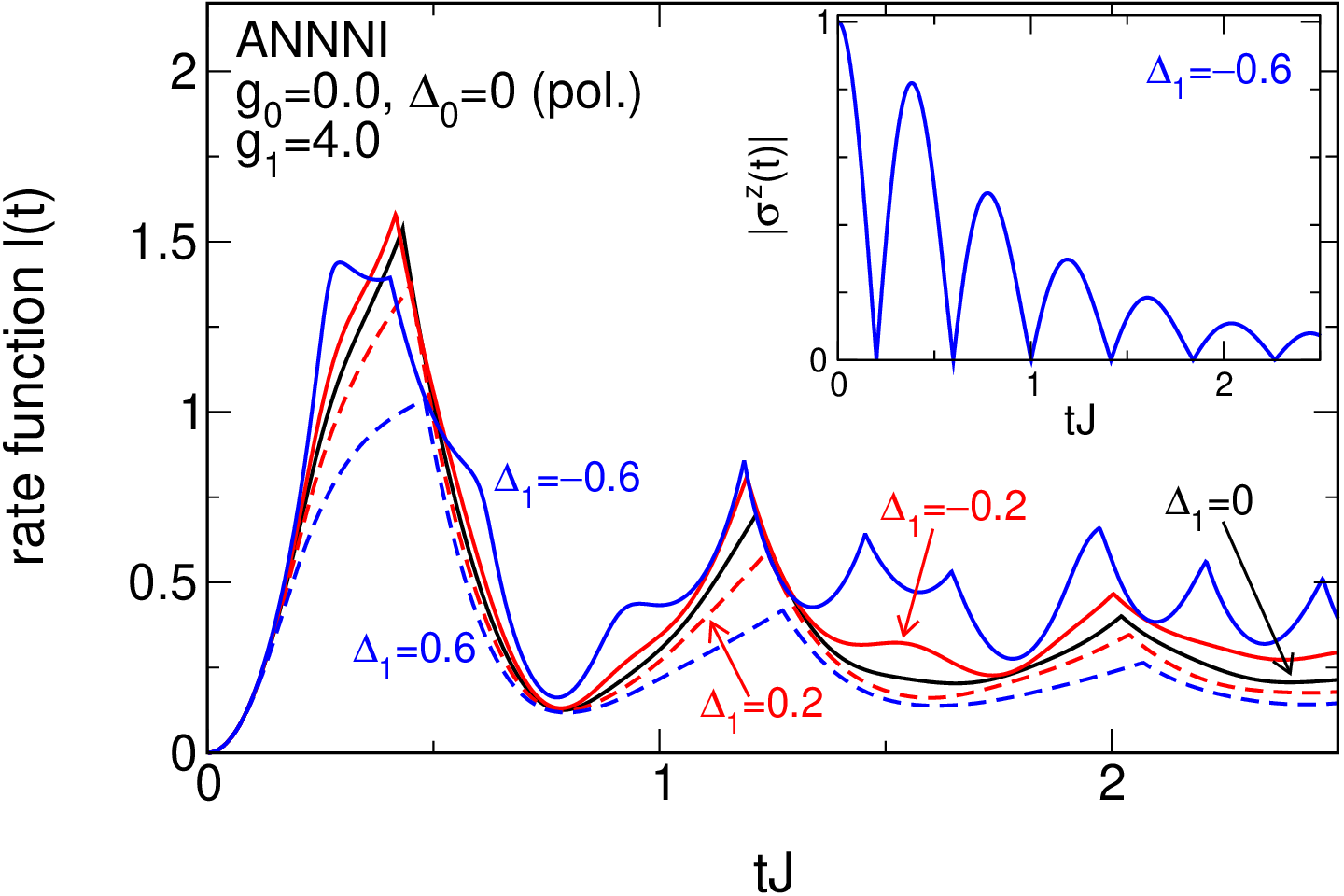}
\caption{(Color online) The same as in Fig.~\ref{fig:ANNNI1} but for a quench from the polarized ground state of the FM phase to the PM phase (dashed-dotted arrow in Fig.~\ref{fig:phasediagramANNNI}). The rate function consistently features non-analyticities. Inset: The order parameter oscillates periodically at sufficiently late times, but the precise connection of its dynamics to $l(t)$ remains elusive.  }
\label{fig:ANNNI2}
\end{figure}

\begin{figure*}[t]
\includegraphics[width=0.475\linewidth,clip]{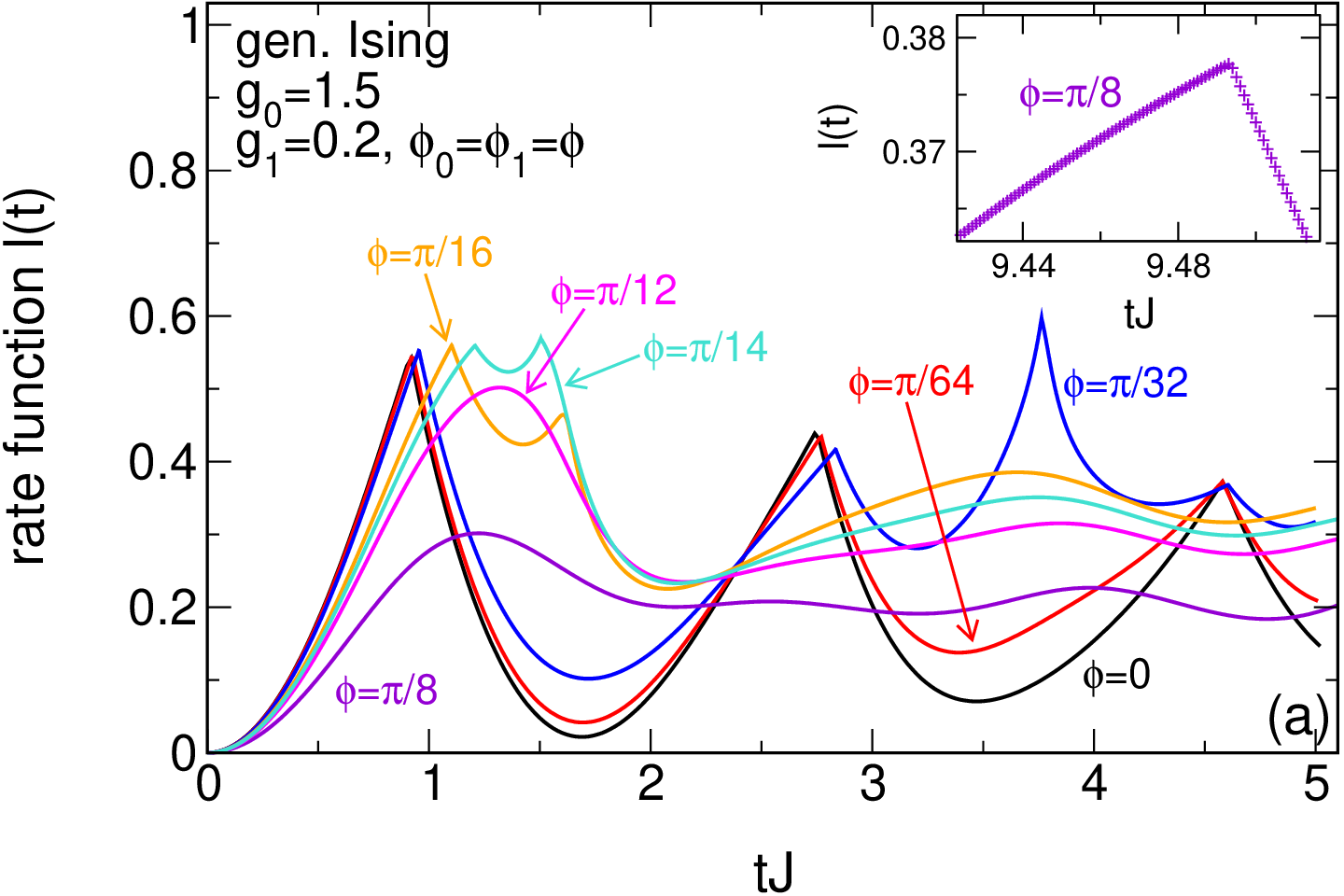}
\hspace*{0.025\linewidth}
\includegraphics[width=0.475\linewidth,clip]{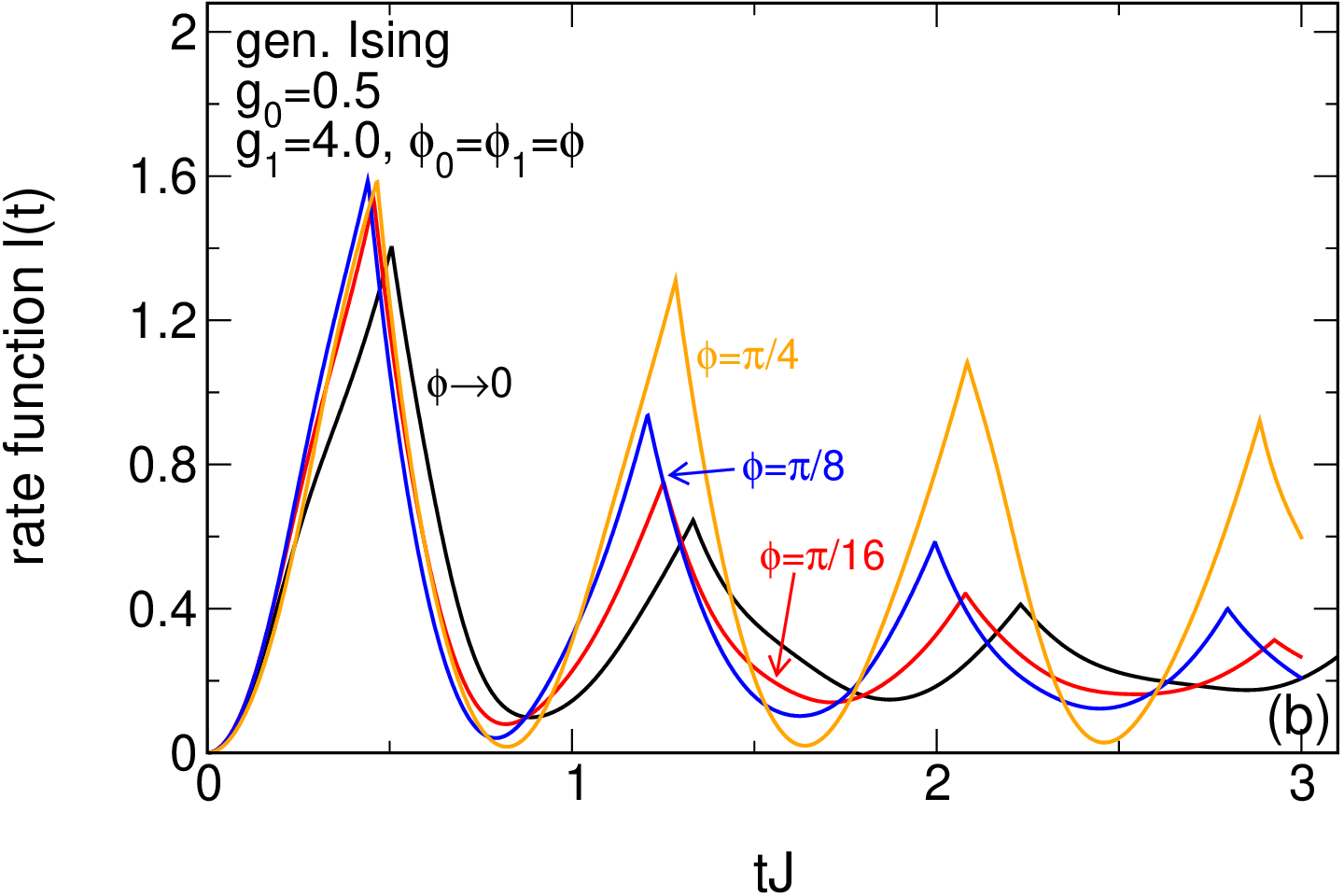}
\caption{(Color online) Rate function for quenches within the generalized Ising model defined in Eq.~\eqref{eq:genIsing}. The model is critical at $g=1$ for all $\phi\geq0$. For $\phi>0$, the model has a unique ground state for any $g$, is not integrable, and the quantum critical point is no longer in the Ising universality class. One consistently observes non-analyticities in the time evolution; they occur periodically only at $\phi=0$. We note that for large angles $\phi$ the non-analyticities appear at rather late times (see the inset) while the behavior for short times is absolutely smooth.} 
\label{fig:genising}
\end{figure*}

We first concentrate on quenches from the PM to the FM phase. Fig.~\ref{fig:ANNNI1}(a) illustrates the effect of successively switching on the `interaction' $\Delta$ for a quench analogous to the one shown in Fig.~\ref{fig:ising1} (it corresponds to the solid arrow in Fig.~\ref{fig:phasediagramANNNI}). The existence of non-analyticities is stable against interactions $\Delta$. Increasing the interaction strength $\Delta$ leads to a decrease of the time scale $t^*$, however, there seems to be no simple quantitative relation. Furthermore,  the critical times $t_n^*$ do not show any periodicity, i.e. it is not possible to write $t_n^*=t^*(\Delta)(n+1/2)$ with some interaction dependent time scale $t^*(\Delta)$ replacing Eq.~\eqref{eq:tstar}. We have furthermore investigated an interaction quench with fixed $g_0=g_1=1.3$ and $\Delta_0=0$, $\Delta_1>0$ (depicted by the dashed arrow in Fig.~\ref{fig:phasediagramANNNI}). The results are shown in Fig.~\ref{fig:ANNNI1}(b). For $\Delta_1=0.2$, one does not leave the PM phase, and the rate function is a smooth function of time. In contrast, for $\Delta_1=0.6$ and $\Delta_1=1$ one enters the FM phase and $l(t)$ becomes non-analytic as expected. Note that the model is strongly non-integrable for those parameters.

Fig.~\ref{fig:ANNNI2} shows DMRG data for the opposite quench from a polarized FM ground state to the PM phase (dashed-dotted arrow in Fig.~\ref{fig:phasediagramANNNI}). The appearance of kinks is again stable against interactions $\Delta\neq0$. Even for the quantum Ising chain the kinks do not occur periodically if one starts from a spin-polarized state; this behavior becomes more pronounced for $\Delta\neq0$. In particular, the evolution between the kinks becomes highly non-trivial including smooth maxima and inflection points, suggesting that for such details interaction effects become important and a simple picture based on the time evolution under a trivial Hamiltonian like $H'$ is not sufficient to describe the dynamics. The order parameter, however, still oscillates periodically at sufficiently late times (see the inset to Fig.~\ref{fig:ANNNI2}). A straightforward connection between order-parameter dynamics and the appearance of non-analyticities in the rate function for the return probability thus remains elusive.

As was pointed out in Ref.~\onlinecite{Heyl}, the non-analytic behavior of the rate function of Eq.~\eqref{eq:returnamplitude} affects the work distribution function of a double quench experiment where one quenches from $H_0$ to $H$  at the time $t=0$ and, at a later time $t$, back to $H_0$. The performed work $W$ possesses the distribution function
\begin{equation}
P(W,t)=\sum_n\big|\bra{n}e^{-\ii Ht}\ket{\Psi_0}\big|^2\,\delta(W-E_n+E_0)~,
\end{equation}
where $\ket{n}$ denote a complete set of eigenstates of the initial Hamiltonian $H_0$ with energies $E_n$. For the work distribution function one can introduce a rate function $r(w,t)$ via
\begin{equation}
P(W,t)=e^{-Lr(w,t)}~,
\end{equation}
with the work density $w=W/L$. Obviously, for vanishing performed work this is identical to the rate function of the return probability, $r(w=0,t)=l(t)$; the non-analytic behavior of $l(t)$ hence manifests in the work distribution function. For the quantum Ising chain it was shown\cite{Heyl} that these non-analyticities at $w=0$ also dominate the behavior of $r(w,t)$ at finite $w>0$, thus making the dynamical phase transition observable in a measurable quantity like the performed work. Given the close similarities of the rate functions $l(t)$ of the quantum Ising chain and the interacting ANNNI model we expect that the dynamical phase transition will also lead to observable effects on the work distribution function in the latter model.

Our results for the ANNNI model show that the appearance of a dynamical phase transition proposed in Ref.~\onlinecite{Heyl} is not an artefact of the quantum Ising chain: the non-analyticities in the time evolution of the rate function for the return probability are stable against integrability-breaking interactions.

\section{Ising model in a tilted field}\label{sec:genIsing}

As last example we consider a generalized Ising chain in a tilted magnetic field.\cite{Pollmann-10,magnfieldIsing} Specifically, we use the parametrization
\begin{equation}
H_\text{gI}=-J\sum_i\bigl[\sigma_i^z\sigma_{i+1}^z+\sigma_i^x+(g-1)(\cos\phi\,\sigma_i^x+\sin\phi\,\sigma_i^z)\bigr]~,
\label{eq:genIsing}
\end{equation}
where $J>0$, $g\geq0$, and $0\le|\phi|\le\pi/2$. For $\phi=0$ we recover the quantum Ising chain \eqref{eq:Ising}. Eq.~\eqref{eq:genIsing} exhibits a quantum phase transition at $g=1$. For any finite angle $\phi$ the ground state is unique and, to the best of our knowledge, the model is not integrable. The transition at $g=1$ is not in the Ising universality class; instead, the correlation length $\xi$ diverges with the exponent $\nu=8/15$. The application of a Jordan-Wigner transformation to the magnetic field term $(g-1)\sin\phi\,\sigma_i^z$ yields a non-local string of fermions. Thus, an analysis of Eq.~\eqref{eq:genIsing} in terms of fermionic degrees of freedom is not possible.

DMRG results for quenches at fixed $\phi_0=\phi_1=\phi$ from $g_0>1$ to $g_1<1$ (which in the limit $\phi\to0$ corresponds to quenches from the PM to the FM phase) and vice versa (polarized FM state to the PM phase) are shown in Fig.~\ref{fig:genising}. The rate function features non-analyticities for arbitrarily large $\phi$. As for the ANNNI model, these non-analyticities do not occur periodically and can potentially be shifted to rather late times. This is particularly striking at $\phi=\pi/8$ in Fig.~\ref{fig:genising}(a) (note the inset) and not an isolated incident (see $\phi=\pi/12$): the time where the first non-analyticity occurs seesms to depend on $\phi$ in a highly non-trivial way.

The dynamics of the generalized Ising model was also investigated by Pollmann \emph{et al.}\cite{Pollmann-10} who studied a \textit{finite-velocity sweep} across the quantum critical point instead of a sudden quench that we focus on. For the continuous sweep a non-integrable perturbation $\phi\neq0$ was found to lead to a broadening of the kinks. A better understanding of the difference between sudden and slow continuous quenches will be the topic of future investigations.

Our results for the generalized Ising model, which is non-integrable and possesses a quantum phase transtition which is not in the Ising universality class, again indicate that the appearance of a dynamical phase transition is a generic feature of sudden quenches across quantum critical points.

\section{Conclusions}\label{sec:conclusion}

In this work we investigated sudden quenches across a quantum critical point of one-dimensional spin models using the time-dependent density matrix renormalization group. We specifically studied the quantum Ising model in presence of two terms which break integrability (next-nearest neighbor interactions and a tilted magnetic field) and showed that in the thermodynamic limit the rate function for the return probability $Ll(t)=-\ln|\bra{\Psi_0}e^{-\ii Ht}\ket{\Psi_0}|^2$ to the initial state $\ket{\Psi_0}$ becomes a non-analytic function of time. This breakdown of short-time expansions is analogous to the failure of high-temperature expansions of the partition function in the vicinity of equilibrium critical points. It was thus termed `dynamical phase transition' and investigated in detail for the integrable (non-interacting) quantum Ising model in Ref.~\onlinecite{Heyl}. The systems studied in our paper are not equivalent to free theories and not integrable. Thus, our results indicate that `dynamical phase transitions' are a generic feature of sudden quenches across a quantum critical points. For quenches originating in an ordered phase we studied the dynamics of the order parameter: it oscillates periodically at sufficiently late times, but a precise connection to the non-analyticities in the return amplitude remains elusive except for the quantum Ising chain.

\acknowledgments

We thank Fabian Essler, Markus Heyl, Stefan Kehrein, Volker Meden, and Giuseppe Mussardo for useful comments and discussions. This work was supported by the DFG via KA 3360-1/1 (CK) and the Emmy-Noether program under SCHU 2333/2-1 (DS) as well as by the Nanostructured Thermoelectrics program of LBNL (CK).

\appendix
\section{Scaling limit in the Ising model}\label{sec:appsl}

For definitness we assume a quench from the PM to the FM phase, i.e. $g_0>1$ and $g_1<1$. In order to study the scaling limit of Eq.~\eqref{eq:tstar} we first introduce the masses
\begin{equation}
M_0=2J(g_0-1)~,\quad M_1=2J(1-g_1)~,
\end{equation}
which are both positive. In terms of the masses the critical time \eqref{eq:tstar} reads
\begin{equation}
t^*=\pi\sqrt{\frac{4J+M_0-M_1}{(M_0+M_1)M_1(4J-M_1)}}~.
\end{equation}
Now taking the scaling limit $J\to\infty$, $g_0,g_1\to 1$, and $a\to 0$ ($a$ is the lattice spacing previously set to one), while keeping the masses and the velocity $v=2Ja$ fixed, we find
\begin{equation}
t^*=\frac{\pi}{\sqrt{(M_0+M_1)M_1}}~.
\label{eq:timeSL}
\end{equation}
The existence of this time scale can also be inferred from the calculation of the boundary partition function Eq.~\eqref{eq:boundarypartitionfunction} directly in the Ising field theory. For a quench across the quantum critical point the derivation requires the introduction\cite{FiorettoMussardo10} of an ultra-violett cut-off in the form of an `extrapolation time' as well as the regularization of infinite-volume divergencies.\cite{LeClair-95,SE12}

We note that the time scale \eqref{eq:timeSL} diverges as~\cite{Heyl} $t^*\propto 1/\sqrt{M_1}$ if one takes the conformal limit of the system after the quench $M_1\to 0$ (i.e. one quenches to the quantum critical point), while quenches away from the quantum critical point ($M_0\to 0$) simply lead to the finite value $t^*=\pi/M_1$.

\section{DMRG calculations}\label{sec:appDMRG}

\begin{figure}[t]
\includegraphics[width=0.95\linewidth,clip]{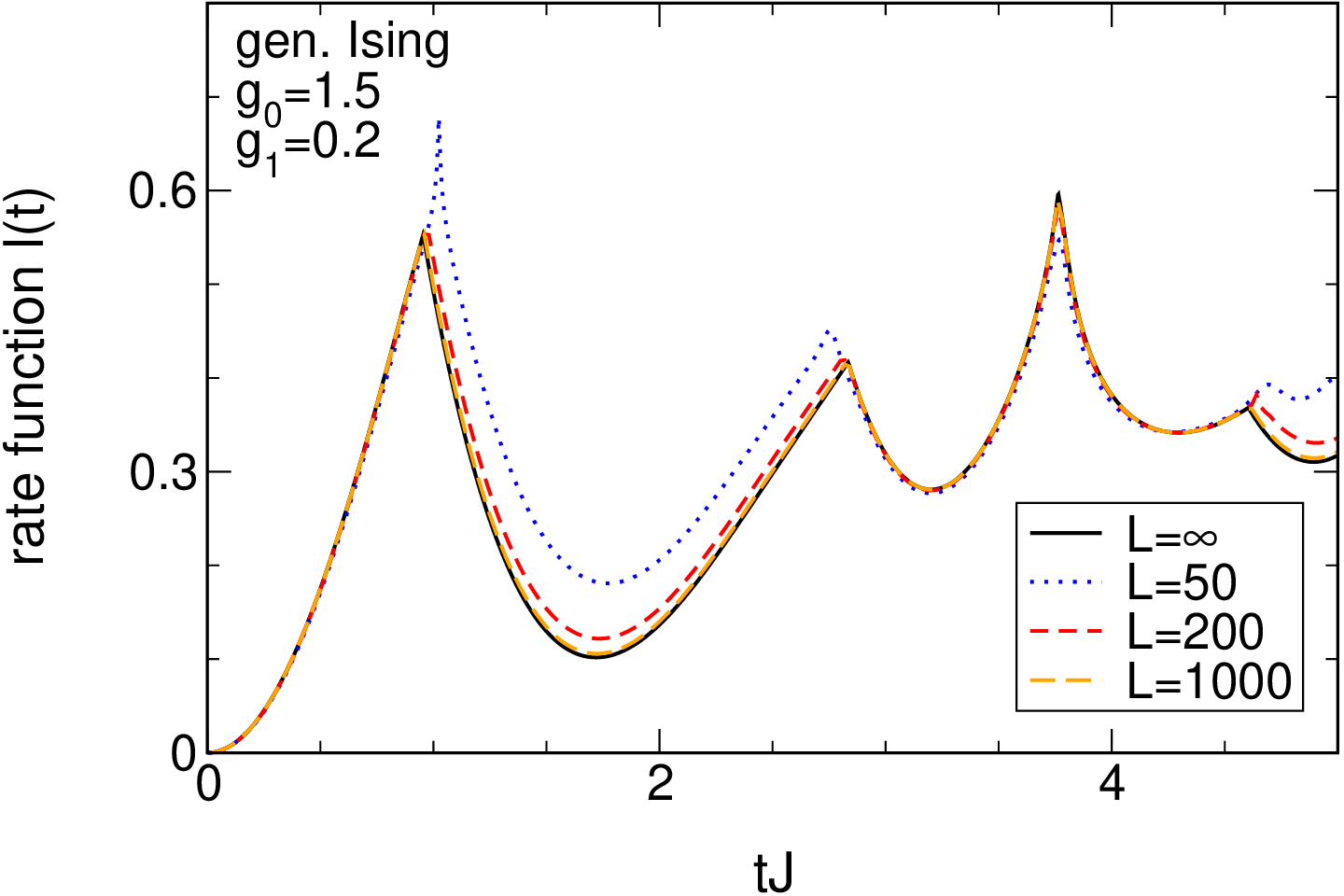}
\caption{(Color online) DMRG calculation of the rate function per site $l(t)$ for a quench across the phase transition of the generalized Ising model of Eq.~(\ref{eq:genIsing}) with $\phi_0=\phi_1=\pi/32$. As the system size $L$ is increased one successively approaches the result obtained directly in the thermodynamic limit $L=\infty$ via an infinite-system DMRG algorithm.\cite{tebd}}
\label{fig:dmrg}
\end{figure}
We study the quench dynamics of the quantum Ising model Eq.~(\ref{eq:Ising}) and its non-integrable generalizations Eqs.~(\ref{eq:ANNNI}) and (\ref{eq:genIsing}) using a DMRG algorithm.\cite{dmrg} The ground state of any one-dimensional system can be expressed in terms of a matrix product state (MPS),\cite{mps}
\begin{equation}\label{eq:mps1}
|\Psi_0\rangle = \sum_{\sigma_n} A^{\sigma_1}A^{\sigma_2}\cdots A^{\sigma_{L}}|\sigma_1\sigma_2\ldots\sigma_{L}\rangle~,
\end{equation}
where $|\vec\sigma\rangle$ denotes the local product basis characterized by $\sigma^z$. In the thermodynamic limit $L\to\infty$ it is convenient to directly work with a translationally-invariant state,\cite{tebd}
\begin{equation}\label{eq:mps2}\begin{split}
|\Psi_0\rangle = \sum_{\vec\sigma} \ldots & \left( A^{[1]{\sigma_{j+1\phantom{+N}}}} A^{[2]{\sigma_{j+2\phantom{+N}}}}\ldots A^{[N]{\sigma_{j+N\phantom{2}}}} \right) \\
\times & \left( A^{[1]{\sigma_{j+1+N}}} A^{[2]{\sigma_{j+2+N}}}\ldots A^{[N]{\sigma_{j+2N}}} \right)\ldots\left|\vec\sigma\right\rangle~.
\end{split}\end{equation}
The matrices $A^{[1\ldots N]\sigma_j}$ are associated with a unit cell of size $N$ (e.g., $N=2$ for the Ising model). We first determine the ground state $|\Psi_0\rangle$ (corresponding to some initial parameters $g_0$, $\Delta_0$, $\phi_0$) by applying an imaginary time evolution $\exp(-\tau H_0)$ to a random state until the energy has converged to ten relevant digits (since we are only considering situations where $H_0$ is gapped, this is very efficient; see below). Thereafter, we carry out the real time evolution $\exp(-\ii H_1t)|\Psi_0\rangle$ using a Hamiltonian $H_1$ which features different parameters $g_1$, $\Delta_1$, $\phi_1$.

In order to implement the above procedure, we employ a standard time-dependent DMRG algorithm.\cite{tdmrg} After factorizing the (real or imaginary) evolution operators $\exp(-c\lambda H)$ using second or fourth order Trotter decomposition, they can be successively applied to Eqs.~(\ref{eq:mps1}) or (\ref{eq:mps2}). At each step $\Delta\lambda$, singular value decompositions are carried out to update (some of the) matrices $A$. We fix their dimension to a constant value $\chi$ which constitutes our main numerical control paramater. The ground state of a gapped system only features finite entanglement and can thus be expressed exactly in terms of a MPS with finite $\chi$ (we are always starting with a gapped $H_0$). During the real-time evolution using $\exp(-\ii H_1t)$, the entanglement increases (we have carried out a global quench). We work with a fixed $\chi$ and for each parameters at hand perform multiple simulations using successively increased $\chi$ (alternatively, one could carry out the ground state search for a given $\chi$ and thereafter fix the discarded weight during the real-time evolution). It turns out that a fairly small $\chi<100$ is sufficient to obtain results which are converged (w.r.t.~increasing $\chi$) on the time scales that we study. For the integrable Ising model, we can exlicitly demonstrate that our data are `numerically exact' by comparing with the analytic solution [see Figs.~\ref{fig:ising1} and \ref{fig:ising2}(b)].

For a finite system described by Eq.~(\ref{eq:mps1}), the return amplitude per site can be computed straightforwardly as the $L$-th root of the overlap between $|\Psi_0\rangle$ and $\exp(-\ii H_1t)\ket{\Psi_0}$. The overlap per site of two infinite MPS [described by matrices $A$ and $\tilde A$ in Eq.~(\ref{eq:mps2})] is given by the $N$-th root of the dominant eignvalue of the transfer matrix
\begin{equation}\begin{split}
T(\{a_1,\tilde a_1\},\{a_N,\tilde a_N\})
 = \sum_{\sigma_1\ldots\sigma_N} & \left(A^{[1]\sigma_1}\ldots A^{[N]\sigma_N} \right)^*_{a_1,a_N} \\
\times &\left(\tilde A^{[1]\sigma_1}\ldots\tilde A^{[N]\sigma_N} \right)_{\tilde a_1,\tilde a_N} .
\end{split}\end{equation}
A comparison between the results obtained by finite- and infinite-system algorithms is shown in Fig.~\ref{fig:dmrg}. 


\end{document}